# A Spin-Resolved Photoemission Study of Photohole Lifetimes in Ferromagnetic Gadolinium

A.V. Fedorov,* T. Valla, F. Liu, P.D. Johnson and M. Weinert

Physics Department, Brookhaven National Laboratory, Upton, NY 11973

P.B. Allen,

Department of Physics and Astronomy, SUNY, Stony Brook NY 11794-3800

## Abstract

High resolution spin-resolved photoemission is used to probe the properties of a Gd(0001) surface state. The state shows both a spin-mixing behavior reflecting the exchange of magnons with the local moments and a reduction of the exchange splitting with increasing temperature. The surface state polarization at low $T$ suggests that the surface layer has an enhanced $T_c$ of 365K or greater. Measurements of the photoemission linewidths show that at low temperatures, the lifetime of a majority spin photohole is predominantly limited by electron-phonon scattering and that of a minority spin photohole by electron-magnon scattering. Since similar behavior may be expected for bulk states close to the Fermi level, the transport properties of this material will also be determined by different decay mechanisms in the two channels.

In photoemission a photon of known energy, $\omega$, is adsorbed and the outgoing electron's energy ($\omega-\varepsilon_k$) and momentum are measured. These properties determine the energy $\varepsilon_k$ and momentum k of the hole state left in the occupied valence bands [1]. Interaction effects, including Coulomb and electron-phonon, cause the sharp line spectrum with shape defined by Im $1/(\omega-\varepsilon_k-i\eta)$ arising from independent electron theory to evolve into Im $1/(\omega-\varepsilon_k-\Sigma(k,\omega))$ where the complex self-energy $\Sigma(k,\omega)$ contains the effects of the interactions. The real part $\Sigma_1(k,\omega) \sim -\omega\lambda_k$ gives a shift in energy and mass enhancement, while the imaginary part $2\Sigma_2(k,\omega) = \Gamma_k(\omega,T)$ gives the lifetime broadening $\hbar/t_k$ where $t_k$ is the typical time before the hole state k scatters into other states k'.

Recent improvements in the energy and momentum resolution of photoemission have enabled studies of $\Gamma_k(\omega,T)$ in two dimensional systems such as surface states, where the k vector has no unresolved component perpendicular to the surface [2]. We have recently reported the first photemission measurement of the mass enhancement $\lambda_k$ for a surface state on Mo [3]. Here we report the first spin-resolved study of $\Sigma(k,\omega)$, by analyzing the spin as well as energy and momentum of photoelectrons emitted from ferromagnetic Gd(0001). Unlike previously studied non-magnetic systems, the hole in Gd scatters from spin waves as well as phonons. Specifically, we demonstrate that at low temperatures the majority spin photohole relaxes mainly by phonon emission whereas the minority spin hole relaxes mainly by spin-wave emission. Further we show that the disappearance of magnetism with increasing temperature is a roughly equal mixture of a (Stoner-like) decrease of spin splitting of conduction bands and an increase of opposite-spin components arising from electron-magnon renormalization. A preliminary report of some aspects of this work has been given elsewhere [4].

The experiments employed a Scienta SES200 electron energy analyzer [4], which can be used either (1) for high-resolution angle-resolved photoemission or (2) for spin-resolved photoemission. Spin polarization is detected with a micro-Mott polarimeter, modified from the design of the Rice University Group [5]. UV photons of energy 21.2 eV were provided by a resonance lamp. The energy resolution in both the angle-resolved and spin-resolved studies is ~ 50 meV. An angular resolution of $\Delta\theta \sim 0.2^o$ is achieved without spin resolution but for spin-resolved studies, because of the reduced intensity, $\Delta\theta \sim \pm 3^o$. Gadolinium films of thickness 200A were evaporated onto a Mo(110) substrate at room temperature and annealed to 750K for one minute to produce well-defined surfaces with a single magnetic domain[6]. Low-energy electron diffraction (LEED) monitored the crystallographic order. The photoemission spectra provided a measure of contaminants, such as oxygen.

Figure 1 shows spectral density maps in the $\Gamma X$ azimuth from the clean Gd(0001) surface at two different temperatures. The width, $\hbar/\tau = -2\,\text{Im}\,\Sigma$, of the surface state at a binding energy of ~ 170meV increases as the temperature is raised from 82K to 300K. The lifetime of the photohole is reduced as a result of increased electron-phonon and electron-magnon scattering at higher T. In the low temperature plot, the surface state has $\hbar/\tau \sim$ constant until the angle of emission exceeds $5^o$. At this point, according to calculated band structures [7], the surface state leaves the bulk band gap and begins to resonate with bulk bands. This accounts for the increased broadening.

Fig. 1 also shows a temperature-dependent shift in the binding energy of the surface state. This has received conflicting interpretations. Nolting et al. [8] show that, depending on the coupling between conduction and localized electrons, the quasiparticle band may show either (1) a Stoner-like reduction of exchange splitting with increasing temperature (weak coupling) or (2), temperature dependent spin-mixing without reduction of exchange splitting (strong coupling). At intermediate coupling, a mixture of the two behaviors is anticipated. Fedorov, *et al.* [9] and Weschke *et al.* [10] both measured a reduction of the exchange splitting with increasing temperature. Similar behavior was seen in temperature-dependent spin-polarized inverse photoemission by Donath *et al.* [11]. In contrast, Li *et al.* [12], using spin-polarized photoemission, saw spin-mixing, rather than a reduction of exchange splitting. In a recent scanning tunneling spectroscopy (STS) study, Getzlaff *et al.* saw a mixture of Stoner and spin-mixing effects [13].

With higher energy resolution than previous spin-resolved photoemission studies [12] we confirm the results found in the STS study.[13] The temperature dependent binding energy shift of the surface-state is interpreted as the thermal reduction of exchange splitting between the occupied majority spin state and an empty minority spin surface state above the Fermi level. The 5d electrons are polarized by their interaction with the localized 4f electrons. As the 4f-derived moments disorder, the exchange splitting reflecting the local mean field decreases and the net polarization in the surface state decreases.

Figure 2 shows spin-polarized photoemission spectra recorded from the surface held at *T* =20K in an angular acceptance $\Delta\theta \sim \pm 3^{o}$. Figure 1 shows that such an angular acceptance adds only minimally to the measured line-widths. The peaks in the majority and minority spin spectra, represented by the filled and open circles respectively, occur at identical binding energies. Earlier experiments [14,15] and calculations [16] indicate that the surface state should be 100% majority spin, due to parallel alignment of the surface and bulk moments. Our observation of a minority spin component could be taken as evidence of either incomplete saturation of the magnetization or the presence of minority domains. However the growth procedure is known to yield a single magnetic domain [6]. We saw no evidence for domains when sampling the spin polarization across the film from one edge to the other. Therefore the coexistence of majority and minority spin components at the same energy is an intrinsic property of the surface state arising from a combination of spin-orbit and spin exchange processes as discussed by Nolting et al [8].

Fitting the spectra with Lorentzians, as in figure 2, shows that the minority spin peak has a larger width than its majority spin counterpart, 116 meV as opposed to 86 meV. Extracting the experimental resolution these widths become approximately 105 meV in the minority spin channel and 70 meV in the majority channel. The lifetime of a photohole can be limited by electron-electron, electron-phonon and electron-impurity scattering, and in a magnetic system, by electron-magnon scattering. At the edge of the surface band, the state also couples to the bulk states as seen in figure 1 for $\theta > 5^\circ$. However such coupling will be stronger in the majority spin channel since the surface state falls closer to the bulk majority-spin band edge than to the bulk minority-spin counterpart [7].

Electron-phonon, electron-magnon and electron-electron scattering each give distinct spin dependent contributions to the scattering rate. Electron-electron scattering by exchange processes favors the two holes in the final state being of opposite spin [17]. From consideration of the total density of states in the spin channels, we estimate the scattering rate from this process to be equal for a majority spin hole and a minority spin hole. The electron-phonon and impurity scattering rate are proportional to the density of states at the hole binding energy for the same spin while the electron-magnon rate is proportional to the density of states for the opposite spin. Since the majority-spin density of states is large while the minority-spin part is small, impurity and electron-phonon scattering should be more important in the majority spin channel. The fact that the minority spin channel is broader indicates an electron-magnon mechanism. At $T=0K$, the minority-spin component of a photohole can scatter to the majority-spin component of a hole state higher in the surface band by emitting a spin wave (tilting the spins of the localized f electrons). The corresponding spin-flip process is not available to the majority-spin component of the photohole at $T=0$ because the localized f spins have saturated magnetization and are not able to tilt upwards when the hole tilts down. At higher temperatures, inelastic scattering can occur back and forth between the two spin channels mediated by the emission or absorption of magnons, but the minority-spin component always has the higher density of final states to scatter into. An approximate treatment [18] using the "$s-f$" Hamiltonian [19] found the result

$$\hbar/t(\downarrow) = \frac{\sqrt{3}}{4} \frac{P'(\uparrow) m^*}{S} \left(\frac{2JSa}{\hbar}\right)^2 \qquad (1)$$

for the decay of the minority $(\downarrow)$ spin component of the surface state due to spin flip scattering with magnon emission. Here J is the $s-f$ exchange parameter giving the exchange splitting 2JS = 0.65 measured [9] for the surface state, m* = 1.21 is the effective mass measured for the surface band, and $P'(\uparrow) = 0.87$ is the experimentally measured majority component of the occupied surface band. With S=7/2 and a = 3.6 A, $\hbar/t(\downarrow) \approx 0.095$ eV. Conversely, replacement of $P'(\uparrow)$ by $P'(\downarrow) = 1 - P'(\uparrow)$ gives $\hbar/t(\uparrow) \approx 0.014$ eV for the majority spin component. Thus at low $T$, the majority spin channel is dominated by electron-phonon scattering whereas the minority spin channel is dominated by electron-magnon scattering. The different spin-dependent lifetimes will clearly be manifested in different spin-dependent mean free paths and transport measurements.

Turning to the temperature dependence of line shapes, first consider electron-electron scattering. As the hole moves to lower binding energies, the phase space for such scattering is reduced. On the contrary, figure 1 shows that as $T$ is raised, the state moves closer to $E_F$ and the peak broadens. This indicates that electron-phonon and electron-magnon scattering play the dominant role and not electron-electron scattering.

Considering electron-phonon coupling alone, the width or inverse lifetime of the photohole is described by the relationship

$$\frac{\hbar}{t}(w,T) = 2p\int_0^\infty dw' a^2 F(w')[1 - f(w-w') + f(w+w') + 2n(w')] \qquad (2)$$

where $f(w)$ and $n(w)$ are Fermi and Bose distribution functions and $\alpha^2F(\omega)$ is the Eliashberg coupling function. For $T \geq \Theta_D/3$ ($\Theta_D$ is the Debye temperature) the phonon contribution to the width is $\hbar/\tau = 2\pi\lambda k_B T$, that is, linear in T. As shown in figure 3, fitting the experimentally determined majority spin line widths with the expression given in Eq. 2, leads to a value of the electron-phonon coupling constant $\lambda \approx 1.0$ in the majority-spin channel. This value may be compared with a value of 1.2 (bulk, spin averaged), extracted from the measured specific heat [20] (using the band-theory density of states and assuming only electron-phonon renormalization) and a theoretical value of 0.4 (also bulk and spin-averaged) obtained in a spin-polarized calculation of the electron-phonon coupling constant [21].

The electron-phonon coupling parameter may be written as $\lambda = N_S \langle I_S^2 \rangle / M \langle \omega^2 \rangle$ where $N_S$ represents the spin-projected density of states at the hole binding energy, $\langle I_S^2 \rangle$ is the Fermi surface average of the electron-phonon matrix element, M is the atomic mass and $\langle \omega^2 \rangle$ is an average phonon frequency. Following the approximate analysis of Skriver and Mertig, [21] but allowing for spin-dependent coupling, one arrives at values $\lambda \approx 0.73$ and 0.31 for the majority and minority spin bulk bands respectively. Our value for the majority spin channel is higher, reflecting perhaps a higher electron density of states in the surface region. Wu *et al*. have calculated an enhanced magnetic moment in the surface layer [7]. Using their calculated majority and minority spin densities in the surface layer, one obtains $\lambda \approx 1.15$ and 0.25 for the surface majority and minority spin electron-phonon coupling, close to the value $\lambda = 1.0$ we find for the majority channel.

The spin-dependent density of states available for scattering is not constant but varies with T as shown in figure 1. A rough estimate suggests that $\lambda$ could change by as much as 10% over the range of T in Fig. 3, decreasing for the majority spin channel and increasing for the minority spin. Electron-magnon scattering will show the opposite behavior, increasing with T in the majority spin channel and decreasing in the minority spin channel. Within the resolution of the experiment and over the limited $T$ range studied, the relative changes compensate, giving the appearance of a constant difference in scattering rates between the two channels.

Accepting that the incomplete polarization in the surface state reflects mixing by magnon scattering between the two channels, it is interesting to examine its $T$ dependence at low $T$. In Gd alloys at such $T$, it has been established that the magnetization follows the Bloch $(T/T_c)^{3/2}$ law reflecting the creation of magnons [22]. In figure 4 we compare the $T$ dependence of the polarization $[(n\uparrow - n\downarrow)/(n\uparrow + n\downarrow)]$ in the surface state with the $T$ dependence of the spin polarization in the background. The slower decline of the surface polarization indicates that the exchange coupling in the surface layer is stronger than in the bulk in agreement with a recent theoretical study by Shick et al. [23]. The background polarization fits the expression $P_o[1-(T/T_c)^{3/2}]$ where $P_o$ represents the "nominal" polarization at 0K and $T_c$ is the bulk Curie $T$ of 290K. This expression does not hold at higher $T$ because of non-linear magnon-magnon scattering effects. Experiments on Fe have shown that at low $T$ both the bulk and surface magnetization follow a $T^{3/2}$ law [24]. Fitting the same expression to our surface-state data requires a $T_c$ of 362 ± 6 K, an indication that the surface may indeed exhibit an enhanced $T_c$. The latter possibility has been discussed in the past, most recently by Tober *et al.* who report a value of 370-380K in general agreement with our value [25], and by Gerion *et al.* who find $T_c \approx 420 - 500 K$ in small clusters [26], higher than our value. However we also note that a recent low energy secondary electron study by Arnold and Pappas concludes that the surface and bulk Tc are identical [27]. The polarization in the present study will also reflect the on-site spin-orbit coupling for the 5d orbitals. A simple model yields a polarization $P = \Delta/\sqrt{(\Delta^2 + z^2)}$ of each quasiparticle state. With a spin-orbit parameter $\xi = 0.3$ eV and exchange splitting $\Delta = 0.7$ eV at 0K, we get a spin-orbit induced mixing R = $(n\downarrow/n\uparrow)$ = (1-P)/(1+P) ~ 5%. This increases to 8% at $T$=150K because of the reduced exchange splitting.

In summary, our spin-resolved photoemission studies with high-energy resolution have allowed us for the first time to study the different contributions to the spin-dependent lifetimes in Gd. We show that at low $T$, the majority spin channel is predominantly limited by electron-phonon scattering whereas the minority spin channel is determined by electron-magnon scattering. As $T$ increases there is an increasing admixture of the two decay modes in each channel. We clearly demonstrate the presence of both spin-mixing behavior and a reduction of the exchange splitting in the Gd(0001) 5d-derived surface state. The polarization in the 5d electrons reflects the degree of magnetic order in the local moments. As the latter disorder, the net polarization in the surface state decreases and the splitting also decreases. Thus the surface state serves as a probe of the local magnetization in the surface region. The surface state polarization at low $T$ suggests that the surface layer has an enhanced $T_c$ of 365K or greater.

The authors acknowledge useful discussions with W. Nolting, G. A. Sawatzky, D. Singh, and R. E. Watson. This work was supported in part by the Department of Energy under contract number DE-AC02-98CH10886. Work of PBA was supported by NSF grant number DMR0089492.

Figure Captions:-

Fig. 1  Upper panel: Spin-integrated photoemitted spectral response for the Gd(0001) surface as a function of binding energy and angle of emission measured from the surface normal. The sample $T$ is 300K and the incident photon energy is 21.2 eV.

Lower panel: As above but now the sample $T$ is 82K.

Fig. 2  Spin-resolved photoemission spectra recorded from the Gd(0001) surface at 20K. The upper and lower spectra represent the emission in the majority and minority spin channels respectively. The lines indicate Lorentzian fits to the spectra superimposed on appropriate backgrounds. The inset shows the relative intensities in the two spin channels.

Fig. 3  The full width half max (FWHM) of the majority spin peak as a function of $T$. The solid line indicates a fit to the data using equation (2) as given in the text.

Fig. 4  The measured polarization in the background (filled diamonds) and surface state (open circles). The lines indicate $T^{3/2}$ fits to the data.

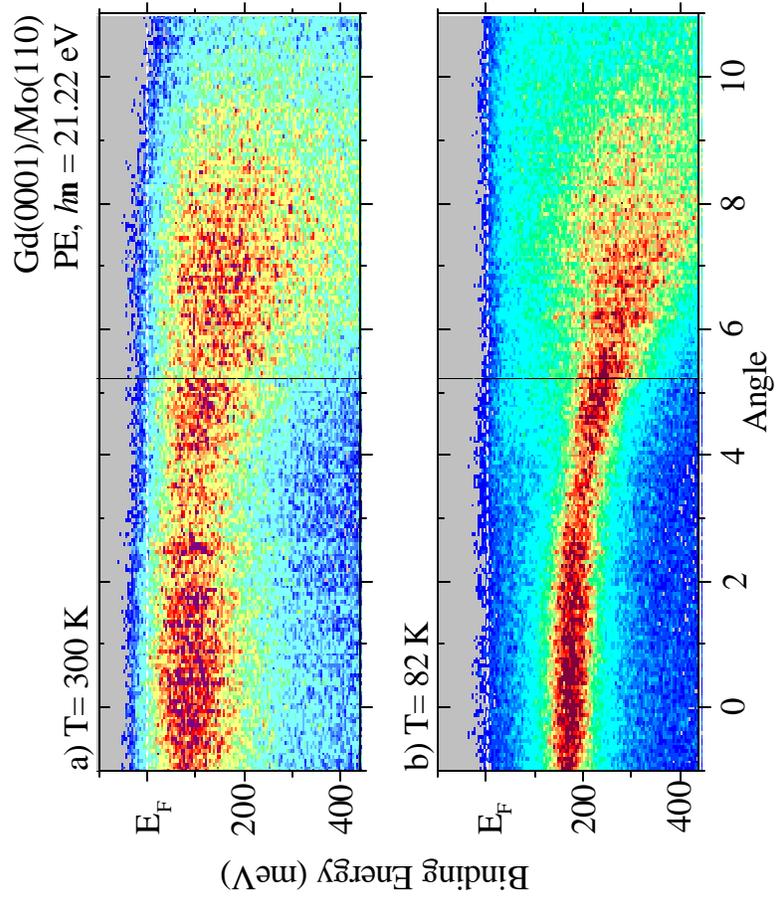

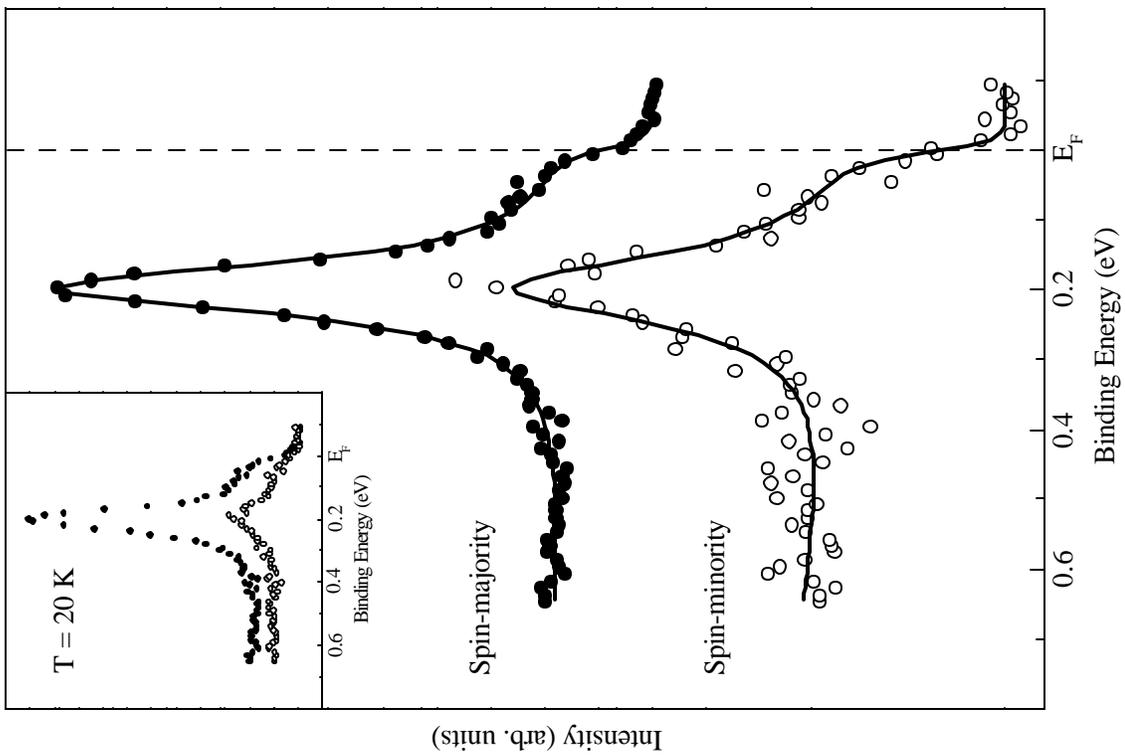

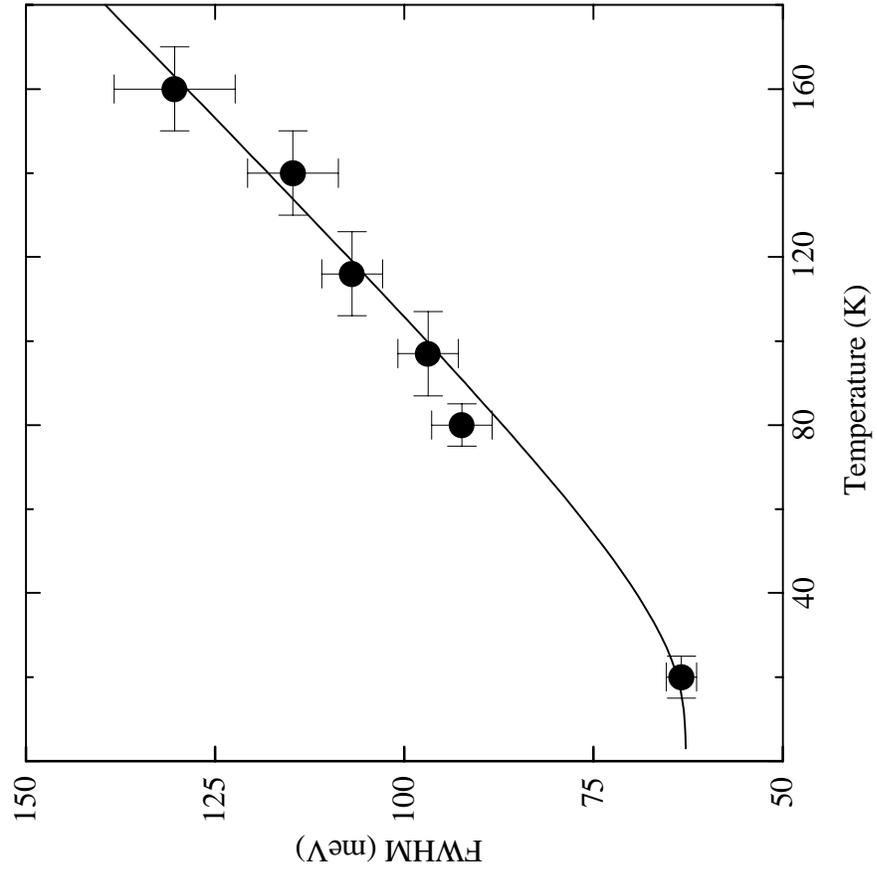

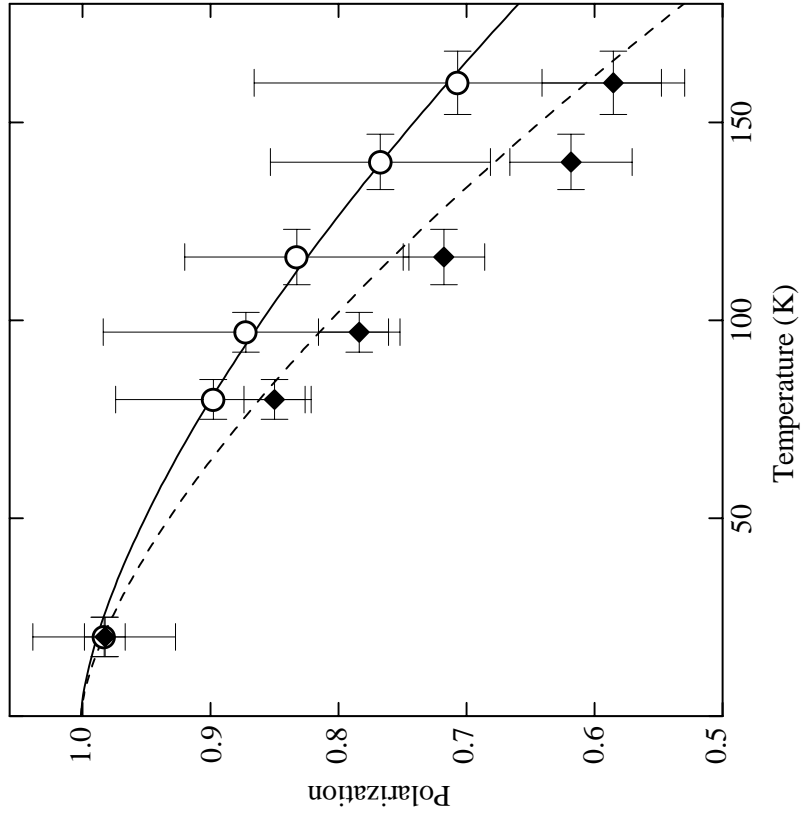